\documentclass[12pt]{article}%
\usepackage[nosort]{cite}
\usepackage{graphicx}
\usepackage{multicol}
\usepackage{amsfonts}
\usepackage{amssymb}
\usepackage{amsmath}
\usepackage{style}
\usepackage[all]{xy}
\usepackage{setspace}
\usepackage{verbatim}
\usepackage{color}
\usepackage{epsfig}
\usepackage[margin=1in]{geometry}
\usepackage{titletoc}
\usepackage{lscape}
\usepackage{tikz}
\usepackage[debug,pageanchor=false]{hyperref}%
\providecommand{\U}[1]{\protect\rule{.1in}{.1in}}
\pdfoutput=1
\numberwithin{equation}{section}

\DeclareMathOperator{\so}{\mathit{so}}
\DeclareMathOperator{\su}{\mathit{su}}

\newcommand\mf{\mathfrak}
\newcommand\ord{\operatorname{ord}}

\def\so{\operatorname{\mathfrak{so}}}

\def\sp{\operatorname{\mathfrak{sp}}}

\def\su{\operatorname{\mathfrak{su}}}

\definecolor{link}{rgb}{.8,.15,.1}
\hypersetup{colorlinks=true,linkcolor=link,citecolor=link,urlcolor=link,linktocpage}
\begin{document}

\date{May 17, 2016}
\title{F-theory  and Unpaired Tensors \\ in 6D SCFTs and LSTs}

\institution{UCSBmath}{\centerline{${}^{1}$Department of Mathematics, University of California Santa Barbara, CA 93106, USA}}

\institution{UCSBphys}{\centerline{${}^{2}$Department of Physics, University of California Santa Barbara, CA 93106, USA}}

\institution{HARVARD}{\centerline{${}^{3}$Jefferson Physical Laboratory, Harvard University, Cambridge, MA 02138, USA}}

\authors{David R. Morrison\worksat{\UCSBmath, \UCSBphys}\footnote{e-mail: {\tt drm@physics.ucsb.edu}},
and Tom Rudelius\worksat{\HARVARD}\footnote{e-mail: {\tt rudelius@physics.harvard.edu}}}

\abstract{We investigate global symmetries for 6D SCFTs and LSTs
having a single ``unpaired'' tensor, that is, a tensor
with no associated  gauge symmetry.
  We verify that 
for every such theory built from F-theory whose tensor has
 Dirac self-pairing equal to $-1$,
the global symmetry algebra is a subalgebra of $\mathfrak{e}_8$.
This result is new if the F-theory presentation of the theory involves
a one-parameter family of nodal or cuspidal rational curves (i.e.,
Kodaira types $I_1$ or $II$) rather than elliptic curves 
(Kodaira type $I_0$).  For such theories, this condition on the
global symmetry algebra appears to fully capture the constraints
on coupling these theories to others in the context of multi-tensor
theories.

We also study the analogous problem for theories whose tensor
has Dirac self-pairing equal to $-2$ and find that the global symmetry algebra
is a subalgebra of $\mathfrak{su}(2)$.  However, in this case there
are additional constraints on F-theory constructions for
coupling these theories to others.
}

\maketitle

%
%
%

\section{Introduction \label{sec:INTRO}}

There have recently been major results in the classification of superconformal field theories (SCFTs) and little string theories (LSTs) in six dimensions  (c.f. \cite{Heckman:2013pva, Gaiotto:2014lca, DelZotto:2014hpa,
Heckman:2014qba, DelZotto:2014fia, Heckman:2015bfa, Bhardwaj:2015xxa, Bhardwaj:2015oru} and earlier works 
\cite{Witten:1995ex, Witten:1995zh, 
WittenSmall, 
Strominger:1995ac,
Ganor:1996mu,
Seiberg:1996vs, 
MorrisonVafaII,
Seiberg:1996qx,
Bershadsky:1996nu, 
Aspinwall:1996vc,
Blum:1997fw, 
Aspinwall:1997ye,
Seiberg:1997zk,
Intriligator:1997dh, 
Brunner:1997gf, 
Hanany:1997gh
}).  
Most notably, \cite{Heckman:2015bfa}, \cite{Bhardwaj:2015xxa}, and \cite{Bhardwaj:2015oru} derived a potentially complete\footnote{In addition
to the theories with unpaired tensors studied here, the two  principal
areas where there is doubt about the completeness of the
classification are the question of frozen singularities
in F-theory \cite{Tachikawa:2015wka}, and the possibility of
theories without a tensor branch.} classification of 6D SCFTs and LSTs using a combination of F-theoretic and field-theoretic constraints.  
The classification is made by studying each theory on its
 ``tensor branch'' (when it has one) where the spectrum of the theory 
consists of the usual supersymmetric multiplets for 6D $\mathcal{N}=(1,0)$
theories:
vector multiplets, tensor multiplets, and hypermultiplets.
The vector multiplets have no scalars, while the scalar in a tensor
multiplet is real and the scalar in a hypermultiplet is quaternionic.

Each simple summand of the gauge algebra is paired with a tensor multiplet
whose scalar expectation value determines the gauge coupling
\cite{Sagnotti:1992qw,DuffMinasianWitten}.  Due to
the requirement of gauge anomaly cancellation, the gauge algebras and
hypermultiplets are severely constrained in this case, and  this is the
part of the recent classification which can be phrased almost completely
in field-theoretic terms.  The ``unpaired tensors'' -- those associated
to no vector -- are under much less control from a field theoretic
perspective and this is where the tools of F-theory have been used
to their fullest.
In this note, we will study global symmetries of the theories with
unpaired tensors, which we hope will be a step towards understanding
them in purely field-theoretic terms.  This extends the work
of \cite{global-symmetries} which analyzed global symmetries for
paired tensors.

Recall that in F-theory, a 6D SCFT or LST is constructed via compactification on an elliptically fibered Calabi-Yau threefold $X \rightarrow B$ with noncompact base $B$.  Simultaneously contracting all of the compact Riemann surfaces within $B$ to zero area yields an SCFT or LST.  On the tensor branch (with some Riemann
surfaces not contracted), the expectation values of the tensors are
given by areas of Riemann surfaces in the SCFT case, and by ratios of such areas
in the LST case.
The data of the threefold $X$ is captured by a Weierstrass equation 
\begin{equation}
y^{2}=x^{3}+f x+g,
\end{equation}
whose discriminant (which vanishes exactly when the fibers are singular)
is given by
\begin{equation}
\Delta \equiv 4f^{3} + 27g^{2}.
\end{equation}
Here $f$, $g$, and $\Delta$ are sections respectively of 
$\mathcal{O}_{B}(-4K_{B})$, $\mathcal{O}(-6K_{B})$,  and 
$\mathcal{O}_{B}(-12K_{B})$, where $K_B$ is the canonical class of $B$. 
Gauge symmetries in F-theory, if present, are determined by the
types of singularity in $X$.

Kodaira classified the types of possible singularities according to the
orders of vanishing of $f$, $g$, and $\Delta$ along the Riemann surface, as shown in Table \ref{Kodairatypetable}.
The cases with no gauge symmetry (the unpaired tensors) are easy to
describe:  either $\Delta$ does not generically vanish along the
Riemann surface (``Kodaira type $I_0$''), or $\Delta$ vanishes to exactly
first order, which implies that $f$ and $g$ cannot generically vanish 
(``Kodaira type $I_1$''), or both $f$ and $g$ vanish with $g$ vanishing
to exactly first order, which implies that $\Delta$ vanishes to exactly
second order (``Kodaira type $II$'').  The type of fiber over the
general point of the Riemann surface is:  a torus for type $I_0$,
a genus zero curve with a node for type $I_1$, and a genus zero
curve with a cusp for type $II$.  In all three cases, the total
space is smooth over the general point of the Riemann surface in
spite of singularities in the fibers themselves.
These are the cases we study in this note.

Our results concern the possible global symmetries which are
explicitly manifest in F-theory.  These global symmetries are associated
to non-compact Riemann surfaces in $B$ on which $\Delta$ vanishes
to order at least two (and to order at least three if $f$ and $g$ also
vanish).  Although the usual rules of F-theory would assign a gauge
algebra to such a Riemann surface, in the scaling limit used to produce
the SCFT or LST the gauge coupling goes to zero and we expect to
find a global symmetry instead.

We study a single compact Riemann surface whose tensor is unpaired,
and which can be contracted to become part of an SCFT or LST.  
The Dirac self-pairing of the corresponding
tensor (which geometrically corresponds to the self-intersection of
the Riemann surface) can be $-1$ or $-2$:  any higher and it would not
be contractible, and any lower would force  pairing with a summand
of the gauge algebra \cite{clusters}.

In the case of self-intersection $-1$, for Kodaira type $I_0$, it was shown in \cite{miranda1986extremal, persson1990configurations,miranda1990persson} that the
global symmetry must be a subgroup of $E_8$, and we confirm that result
with a few more details here.  Our first new result is that the same
statement -- the global symmetry is a subgroup of $E_8$ -- also
holds for Kodaira types $I_1$ and $II$.

In the case of self-intersection $-2$, the situation is somewhat 
different.  It is easy to see that for Kodaira type $I_0$ there
can be no global symmetry (of geometric origin).  However, our
second main result is a classification of the things which can
occur for Kodaira types $I_1$ and $II$ in this case.
In every case, the global symmetry is a subgroup of $SU(2)$.

The classification techniques of \cite{Heckman:2015bfa} show that the ``subgroup of
$E_8$'' criterion for self-intersection $-1$
is a very powerful one, essentially allowing the
completion of the classification.  However, the corresponding
``subgroup of $SU(2)$'' criterion for self-intersection $-2$ does
not seem to capture all of the information from F-theory.  We give 
examples and discussion of this issue in section \ref{sec:UNUSUAL}.

There is some overlap between this paper and  concurrent work
of Johnson and Taylor \cite{JohnsonTaylor}.

\section{The Tools}

The base encodes the structure of the tensor branch: each compact curve of self-intersection $-1$ or below in the base has an associated tensor multiplet, with the size of the curve giving the vev of the scalar in that multiplet.  The Dirac pairing on the string charge lattice of the SCFT or LST is specified by the intersection matrix of the base.  For instance, a base with a curves of self-intersection $-3$ and a curve of self-intersection $-2$ intersecting at a single point would have intersection matrix:
\begin{equation}
A_{IJ} = \left( \begin{array}{cc} -3 & 1\\ 1 & -2  \end{array} \right).
\end{equation} 
We use the shorthand
\begin{equation}
3 \,\, 2
\end{equation}
to illustrate this configuration of curves/tensor multiplets.  The numbers indicate the negatives of the self-intersection numbers of the curves in the base, with adjacent curves intersecting at precisely one point.  An SCFT is characterized by a base of negative definite intersection form, whereas an LST is characterized by a base of negative semidefinite intersection form whose kernel is of rank 1.

\begin{table}[t]
\begin{center}
\begin{tabular}{ccc|c|c|c}
ord($f$)	&ord($g$)		&ord($\Delta$)		&type			&singularity			&non-abelian algebra \\
\hline
$\geq0$	&$\geq0$		&0				&I${}_0$			&none				& none \\
$0$		&$0$			&1				&I${}_1$			&none				& none \\
$0$		&$0$			&$n\geq2$		&I${}_n$			&$A_{n-1}$			&$\su(n)$ or $\sp([n/2])$ \\
$\geq1$	&$1$			&2				&II				&none				& none \\
$1$		&$\geq2$		&3				&III				&$A_1$				&$\su(2)$\\
$\geq2$	&$2$			&4				&IV				&$A_2$				&$\su(3)$ or $\su(2)$ \\
$\geq2$	&$\geq3$		&6				&I${}_0^\ast$		&$D_4$				&$\so(8)$ or $\so(7)$ or $\mathfrak{g}_2$ \\
$2$		&$3$			&$n\geq7$		&I${}_{n-6}^\ast$	&$D_{n-2}$			&$\so(2n-4)$ or $\so(2n-5)$\\
$\geq3$	&$4$			&8				&IV${}^\ast$		&${E}_6$		&$\mathfrak{e}_6$ or $\mathfrak{f}_4$ \\
$3$		&$\geq5$		&9				&III${}^\ast$		&${E}_7$		&$\mathfrak{e}_7$ \\
$\geq4$	&$5$			&10				&II${}^\ast$		&${E}_8$		&$\mathfrak{e}_8$ \\
$\geq4$	&$\geq6$		&$\geq12$		&non-minimal		&-					&- 
\end{tabular}
\end{center}
\caption{Singularity types with associated non-abelian algebras.}
\label{Kodairatypetable}
\end{table}

The Kodaira type of the fiber above a compact curve in the base specifies the gauge algebra that is paired with the tensor multiplet associated to that curve.  In many cases, the degrees of vanishing of $f$, $g$, and $\Delta$ suffice to determine the gauge algebra exactly.  However, in other cases, information regarding monodromy of the fiber is needed to determine the gauge algebra.  The relevant data are summarized in Table \ref{monodromycovertable}.  In all cases except that of $I_0^*$, the monodromy cover is of the form $\psi^2-X$, and the cover splits if and only if $X$ is a perfect square.  For $I_m$, $IV$, $I_n^*$, and $IV^*$, a split cover gives rise to gauge algebra $\mathfrak{su}(m)$, $\mathfrak{su}(3)$, $\mathfrak{so}(2n+8)$, and $\mathfrak{e}_6$, respectively.  A non-split cover gives rise to $\sp([n/2])$, $\su(2)$, $\mathfrak{so}(2n+7)$, and $\mathfrak{f}_4$, respectively.  For $I_0^*$, the monodromy cover may be irreducible or it may split into two or three components, giving rise to $\mathfrak{g}_2$, $\mathfrak{so}(7)$, or $\mathfrak{so}(8)$, respectively.

\begin{table}[t]
\begin{center}
\begin{tabular}{|c|c|} \hline
type 
&equation of monodromy cover\\ \hline\hline
I${}_m$, $m\ge3$
&$\psi^2+(9g/2f)|_{z=0}$
\\ \hline 
IV
&$\psi^2-(g/z^2)|_{z=0}$
\\ \hline
I${}_0^*$
& $\psi^3+(f/z^2)|_{z=0}\cdot\psi+(g/z^3)|_{z=0}$
\\ \hline
I${}_{2n-5}^*$, $n\ge3$
&$\psi^2+\frac14(\Delta/z^{2n+1})(2zf/9g)^3|_{z=0}$
\\ \hline 
I${}_{2n-4}^*$, $n\ge3$
&$\psi^2+(\Delta/z^{2n+2})(2zf/9g)^2|_{z=0}$
\\ \hline 
IV${}^*$
& $\psi^2-(g/z^4)|_{z=0}$
\\ \hline 
\end{tabular}
\end{center}
\caption{Monodromy covers for Kodaira fiber types.}
\label{monodromycovertable}
\end{table}

Curves of self-intersection $-3$ or below necessarily have degenerate fibers which correspond to non-Abelian gauge algebras.  Thus, the tensor multiplets associated with these curves are always ``paired."  On the other hand, curves of self-intersection $-1$ or $-2$ can support fibers that do not produce non-Abelian gauge algebras.  Specifically, when a $-1$ curve or $-2$ curve has a fiber of Kodaira type $I_0$, $I_1$, or $II$, it will not be paired with a gauge algebra and is referred to as an ``unpaired tensor."

In \cite{Heckman:2013pva}, the ``gauging condition" for unpaired $-1$ curves was introduced.  Given a configuration of curves:
\begin{equation}
\overset{\mathfrak{g}_L}L \,\, 1  \,\, \overset{\mathfrak{g}_R}R,
\end{equation}
where $L$ and $R$ represent some curves holding gauge algebras $\mathfrak{g}_L$ and $\mathfrak{g}_R$, respectively, the ``gauging condition" imposes the constraint $\mathfrak{g_L} \oplus \mathfrak{g}_R \subset \mathfrak{e}_8$.  However, up to this point, this gauging condition has only been verified explicitly in F-theory for $-1$ curves of Kodaira fiber type $I_0$.  In principle, one could imagine that $-1$ curves of fiber types $I_1$ or $II$ might give rise to distinct unpaired tensors, and might even yield different gauging conditions than the $\mathfrak{e}_8$ condition for type $I_0$ fibers.  One of our main results in this note is that this is \emph{not} true: $-1$ curves of fiber type $I_0$, $I_1$, and $II$ all satisfy the gauging condition $\mathfrak{g_L} \oplus \mathfrak{g}_R \subset \mathfrak{e}_8$ in F-theory.  In fact, we have no evidence that these different Kodaira types give rise to different 6D SCFTs, nor do we have any evidence that any field-theoretic constraints stronger than the $\mathfrak{e}_8$ gauging condition are needed to match F-theoretic constraints involving unpaired $-1$ tensors.\footnote{A more general challenge is to determine
which geometric features are needed for distinuishing among physical
theories, and which ones are not relevant.  For example, the most
famous of the unpaired tensors, the ``E-string theory,'' is described
by a Weierstrass equation in which $f$ has multiplicity $4$ at a point
and $g$ has multiplicty $6$.  The multiplicities matter, but the precise
polynomials used to construct the model do not matter to the physical theroy.}  We are led to suspect that this gauging condition is the one and only field-theoretic constraint involving unpaired $-1$ tensors in 6D SCFTs and LSTs.

Unpaired $-2$ tensors, on the other hand, are considerably more convoluted in F-theory.  Here, as for $-1$ tensors, F-theory suggests the presence of a ``gauging condition": given a set of curves $\Sigma_i$ carrying gauge algebras $\mathfrak{g}_i$ touching an unpaired $-2$ curve,
 the ``gauging condition" imposes the constraint $\bigoplus \mathfrak{g_i} \subset \mathfrak{su}_2$.  However, this gauging condition is clearly not the only constraint seen in F-theory.  Firstly, a $-2$ curve of type $I_0$ cannot meet any curve carrying a degenerate fiber, so $\bigoplus \mathfrak{g_i}$ is trivial  for an $I_0$ curve.  Furthermore, as we will show in section \ref{sec:UNUSUAL}, there are additional theories which obey this $\mathfrak{su}_2$ gauging condition but which cannot be realized in F-theory.  We will further identify a field-theoretic explanation for the nonexistence of these theories, but we will see that this raises additional questions from a field theory perspective.

Our analysis proceeds by constraining residual orders of vanishing of $f$, $g$, and $\Delta$.  Given a curve $\Sigma = \{ z = 0 \}$, we define $a$, $b$, and $d$ to be the order of vanishing of $f$, $g$, and $\Delta$, respectively, along $\Sigma$.  Next, we define
\begin{equation}
\tilde f = \frac{f}{z^a}\,,~~~~\tilde g = \frac{g}{z^b}\,,~~~~\tilde \Delta = \frac{\Delta}{z^d}.
\end{equation}
These are sections of $\mathcal{O}(-4 K_B-a \Sigma)$, $\mathcal{O}(-6 K_B-b \Sigma)$, and $\mathcal{O}(-12 K_B-d \Sigma)$, respectively.  We define residual vanishings on $\Sigma$ by
\begin{align}
\tilde a &= (-4 K_B-a \Sigma) \cdot \Sigma  = -4 (m-2) + m a \nonumber \\
\tilde b &=    (-6 K_B-b \Sigma) \cdot \Sigma  = -6 (m-2) + m b     \nonumber \\ 
\tilde d &=  (-12 K_B-d \Sigma) \cdot \Sigma  = -12 (m-2) + m d  \nonumber
\end{align}
Here, $\Sigma \cdot \Sigma = -m$ is the self-intersection number of $\Sigma$.  Suppose now that $\Sigma$ intersects each of a collection of curves $\Sigma_k'$ at respective points $P_k$.  We then define
\begin{equation}
\tilde{a}_{P_k} = \ord _{P_k} \tilde{f} \,,~~~~ \tilde{b}_{P_k} = \ord _{P_k}\tilde{g}\,,~~~~\tilde{d}_{P_k} = \ord_{P_k} \tilde{\Delta}.
\end{equation}
If $f$, $g$, and $\Delta$ vanish respectively to order $a_k'$, $b_k'$, and $d_k'$ along $\Sigma_k'$, we always have $\tilde{a}_{P_k} \geq {a}_{k}'$, $\tilde{b}_{P_k}  \geq {b}_{k}'$, $\tilde{d}_{P_k} \geq  {d}_{k}'$.  The precise relations between these values depends on the fiber types of the intersecting curves.  The cases in which $\Sigma$ supports of fiber of type $I_n, n \geq 2$, $I_0^*$, $III$, $IV$, $IV^*$, $III^*$, or $II^*$ were worked out in \cite{global-symmetries}.  The remaining cases of $I_0$, $I_1$, and $II$--the fiber types that give rise to unpaired tensors--will be discussed in section \ref{sec:UNPAIRED}.

The residual orders of vanishing must satisfy
\begin{equation}
\tilde{a} \geq \sum_k \tilde{a}_{P_k} \,,~~~~ \tilde{b} \geq \sum_k \tilde{b}_{P_k}\,,~~~~\tilde{d} \geq \sum_k \tilde{d}_{P_k}.
\end{equation}
This condition tightly constrains the allowed fiber types that are allowed to intersect, and it plays the key role in the analysis that follows.

\section{Adjacencies for Unpaired Tensors}\label{sec:UNPAIRED}

An unpaired tensor in an SCFT or an LST constructed using F-theory may only meet tensors carrying particular gauge groups.  In particular, we show in this section using the methods of \cite{global-symmetries} that an empty $-1$ curve can only meet curves $\Sigma_i$ carrying gauge algebras $\mathfrak{g}_i$ if $\bigoplus_i \mathfrak{g}_i \subset \mathfrak{e}_8$.  Similarly, an empty $-2$ curve can only meet curves $\Sigma_i$ carrying gauge algebras $\mathfrak{g}_i$ if $\bigoplus_i \mathfrak{g}_i \subset \mathfrak{su}_2$.  We show that these results hold regardless of whether the unpaired tensor has Kodaira fiber type $I_0$, $I_1$, or $II$.

\subsection{Type $I_0$}

A compact curve of fiber type $I_0$ or type $I_1$ may have self-intersection $-1$ or $-2$.  In the latter case, we have residual orders of vanishing $\tilde{a} = \tilde{b} = \tilde{d}=0$, which means that no curves with singular fibers can intersect a $-2$ curve of type $I_0$.  In the former case, the classification was carried out by Miranda and Persson in 
\cite{miranda1986extremal,persson1990configurations,miranda1990persson}.  We briefly summarize the analysis here.

A $-1$ curve of type $I_0$ has degrees of vanishing $(\tilde a, \tilde b, \tilde d) = (4,6,12)$.  The direct sum of the sublattices of the Picard group generated by components of fibers that don't meet the section forms a sublattice of rank equal to the sum of the ranks of the fibers, which is $\leq 8$.  This implies that the sum of the ranks of the singular fibers on curves intersecting this $I_0$ curve is less than or equal to 8.  Furthermore, if it is exactly 8, then the the discriminant factorizes as a perfect square.  Additional restrictions come from considering the Jacobi $j$-function.

The values of $\tilde a_P, \tilde b_P, \tilde d_P$ and the rank $r$ for each of the fiber types intersecting an $I_0$ are simply those shown in Table \ref{Kodairatypetable}.   The full list of collisions can be found in \cite{persson1990configurations}, but we content ourselves with a table of the maximal symmetry algebras, shown in Table \ref{typeI0tableb}.

\begin{table}
\begin{center}
\begin{tabular}{c|c} 
Kodaira Types & Symmetry Algebras \\\hline
$II^*$ & $\mf{e}_8 $ \\
$III^* \oplus III$ &$\mf{e}_7 \oplus \mf{su}(2) $ \\
$IV^* \oplus IV$ & $\mf{e}_6 \oplus  \mf{su}(3) $ \\
$I_9$ & $\mf{su}(9)  $ \\
$I_4^*$ &$\mf{so}(16)  $ \\
$I_0^* \oplus I_0^*$ & $\mf{so}(8) \oplus \mf{so}(8)  $ \\
$I_2^* \oplus I_2$& $\mf{so}(12) \oplus \mf{su}(2)  $ \\
$I_1^* \oplus I_4$ & $\mf{so}(10) \oplus \mf{su}(4)  $ \\
\end{tabular}
\caption{Maximal type $I_0$ intersections for $-1$ curves.}
\label{typeI0tableb}
\end{center}
\end{table}

\subsection{Type $I_1$}

For the case of a curve $\{ z= 0\}$ with $I_1$ fiber, we consider the most general forms of $f$, $g$, and $\Delta$:
\begin{align}
f & = -\frac{1}{48} \phi^2 + f_1 z + O(z^2) \nonumber \\
g &= \frac{1}{864} \phi^3 + g_1 z + O(z^2)   \label{i1form} \\
\Delta &=  \frac{1}{192} \phi^3 ( 12 g_1 + \phi f_1 ) z + O(z^2) \nonumber
\end{align} 
We see that this curve cannot meet a curve $\{ \sigma = 0 \}$ of type $II^*$ or $III^*$.  In any such case, we would necessarily have $\sigma^2 | \phi$ and would introduce a $(4, 6, 12)$ point.

We next note that this curve can only meet a $IV^*$ curve that is non-split.  For such an intersection, we must have $\sigma^2 | \phi$, $\sigma^3 | f_1$, $\sigma^4 | g_1$.  The resulting point of intersection    then has orders of vanishing of $f$, $g$, and $\Delta$ given by $4$, $5$, and $10$, respectively, and the residual vanishings at the point are given by $(\tilde{a}_P, \tilde{b}_P, \tilde{d}_P) = ( 4,6,10)$.  The monodromy cover of $\{ \sigma = 0 \}$ is given by $\psi^2 - \frac{g}{\sigma^4} |_{\sigma=0}$, and $ \frac{g}{\sigma^4} |_{\sigma=0} = \frac{g_1}{\sigma^4}|_{\sigma=0} z + f_2  \frac{g_2}{\sigma^4} |_{\sigma=0} z^2 + O(z^3)$.  This can only be a perfect square if  $\frac{g_1}{\sigma^4}|_{\sigma=0} = 0$, but this would in turn raise the order of vanishing of the intersection point to $(4,6,12)$.  Thus, this $IV^*$ curve can only be non-split.

We next consider intersections with curves of type $I_n$.  We define $\Delta_R$ via
\begin{equation} 
 \frac{1}{192}\phi^3 \Delta_R = \tilde{\Delta} |_{z=0}=  \frac{1}{192}\phi^3 ( 12 g_1 + \phi f_1 ). 
\end{equation}
We necessarily have $\sigma^n | \Delta_R$.  From the form (\ref{i1form}), we note that $\tilde{a} = 2 \text{ deg} \phi$, $\tilde{b} = 3 \text{ deg} \phi$, yielding $\text{deg} \phi = 2$.  As a result, we have $\tilde{\Delta} = \text{deg} \tilde \Delta = 3 \text{ deg}\phi + \text{deg} \Delta_R$, which means $\text{deg} \Delta_R =\text{deg} \tilde{\Delta} - 6$.  Since $\sigma^n | \Delta_R$, we conclude that $n \leq \text{deg} \tilde{\Delta} - 6$.  As far as we can tell, this $I_n$ curve may be split or non-split.

A very similar analysis applies to the case of $I_n^*$ curves meeting a type $I_1$ curve.  Once again, we have $ \text{deg} \tilde \Delta = 3 \text{ deg}\phi + \text{deg} \Delta_R$ and thus $\text{deg} \Delta_R = \text{deg} \tilde{\Delta} - 6$.  However, we now have $\sigma^1 | \phi$, $\sigma^2 \nmid \phi$, which means $\sigma^{n+3} | \Delta_R$.  We therefore have $n \leq \text{deg} \tilde{\Delta} - 9$.  Once again, this $I_n^*$ curve may be split or non-split as far as we can tell at first glance.  However, in the split case, we minimally have $\tilde{d}_P = n+7$ rather than $\tilde{d}_P = n+6$.  The monodromy cover splits only if $(\Delta / \sigma^{n+2})(2\sigma f/9g)^3|_{\sigma=0}$ is a perfect square for $n$ odd and only if $(\Delta / \sigma^{n+2})(2\sigma f/9g)^2|_{\sigma=0}$ is a perfect square for $n$ even.  Note that $(2\sigma f/9g)|_{\sigma=0} \sim z^0 + O(z^1)$ for an intersection with $I_1$ curve $\{ z=0\}$, while $(\Delta / \sigma^{n+2}) |_{\sigma=0} = O(z^1)$.  The only way this can be a perfect square is if the coefficient of $z^1$ in $\Delta$ is $O(\sigma^{n+7})$, in which case $(\Delta / \sigma^{n+2})(2\sigma f/9g)^3|_{\sigma=0}$ vanishes and the monodromy cover splits.

The splitting of $I_0^*$ is rather non-trivial in this context.  The monodromy cover for $I_0^*$ takes the form,
\begin{equation}
\psi^3 + (f/\sigma^2)|_{\sigma=0} \psi + (g/\sigma^3)|_{\sigma=0}
\label{i0monodromy}
\end{equation}
To get a fully split $I_0^*$ fiber with $\mf{so}(8)$ symmetry algebra, this cover must factorize as
\begin{equation}
(\psi + \alpha) (\psi + \beta) ( \psi -\alpha -\beta).
\end{equation}
We claim that the degree of vanishing is $\tilde d_{P} \geq 7$ i.e. it is greater than the non-split case of $\tilde d_P = 6$.  To see this, we expand $\alpha$ and $\beta$ order by order in $z$, $\alpha= \alpha_0 + \alpha_1 z + O(z^2)$, $\beta = \beta_0 + \beta_1 z + O(z^2)$.  Then, comparing the split monodromy cover to the general one of (\ref{i0monodromy}) form (\ref{i1form}), we have
\begin{align}
- \frac{1}{48} \phi^2 &= -\alpha_0^2 - \beta_0^2 - \alpha_0 \beta_0+ O(\sigma)  \\
 \frac{1}{864} \phi^3 &= \alpha_0 \beta_0 ( \alpha_0 + \beta_0) + O(\sigma)
\end{align}
Solving these equations for $\alpha_0$ in terms of $\beta_0$ gives either $\alpha_0 = \beta_0 $ or $\alpha_0 = -2\beta_0$.  Next, working to first order, we have
\begin{align}
(f_1 /\sigma^2)|_{\sigma=0} &= -2 \alpha_0 \alpha_1 - 2 \beta_0 \beta_1 - \alpha_0 \beta_1 - \alpha_1 \beta_0 \\
(g_1/ \sigma^3)|_{\sigma=0}  &= 2 \alpha_0 \alpha_1 \beta_0 + \alpha_0^2 \beta_1 + \alpha_1 \beta_0^2 + 2 \alpha_0 \beta_0 \beta_1 + O(\sigma)
\end{align}
The last step is to plug $f_1$, $g_1$, and $\phi$ into $\Delta = \frac{1}{192} \phi^3 (12 g_1 + \phi f_1) z + O(z^2)$.  We find that indeed, for $\alpha_0 = \beta_0 $ or $\alpha_0 = -2 \beta_0$, we have $\Delta = z O(\sigma^7) + O(z^2)$.  Thus, $\tilde{\Delta}_P \geq 7$.

It is always possible to write the monodromy cover of a $I_0^*$ curve meeting our $I_1$ curve in a semi-split form provided this $I_0^*$ curve does not intersect any other fiber types that lead to a splitting.  Thus, an $I_0^{*,ss}$ curve intersects an $I_1$ curve with $(\tilde a_P, \tilde b_P, \tilde d_P) = (2,3,6)$.

Finally, we consider the remaining cases of type $II$, $III$, and $IV^{ns}$ fibers.  The analysis here is straightforward, and we find that $\sigma^m | \Delta_R$, with $m=1, 2, 2$ for $II$, $III$, $IV^{ns}$, respectively.  Furthermore, $(\tilde{a}_P, \tilde{b}_P, \tilde{d}_P) =$ $(2,3,4)$, $(2,3,5)$, and $(2,3,5)$ for these three respective cases.  

If the type $IV$ fiber is split, we must have $g_1 = 0$.  This in turn implies $\sigma^3 | \Delta_R$, and $(\tilde{a}_P, \tilde{b}_P, \tilde{d}_P) = (2,3,6)$.

We summarize the above possibilities in Table \ref{typeI1table}.

\begin{table}
\begin{center}
\begin{tabular}{c|c|c|c|c} 
Fiber Type & $\tilde{a}_P$ & $\tilde{b}_P$ & $\tilde{d}_P$  & $\text{ord}_P \Delta_R$ \\\hline
$IV^{*,ns}$ & 4&6 &10 & 4 \\
$I_n^{*,s}, n\geq 1$ & 2&3 &n+7 &n+4 \\
$I_n^{*,ns}, n\geq 1$ & 2&3 &n+6 &n+3 \\
$I_0^{*,s}$ & 2&3 &7 &4 \\
$I_0^{*,ss}$ & 2&3 &6 &4 \\
$I_n$& 0 & 0 & n+3 & n \\ 
$IV^{s}$ &2 &3 &6 & 3 \\ 
$IV^{ns}$ &2 &3 &5 & 2 \\ 
$III$ &2 &3 &5 & 2 \\ 
$II$ &2 &3 &4 & 1 \\ 
\end{tabular}
\caption{Orders of vanishing at type $I_1$ intersections.}
\label{typeI1table}
\end{center}
\end{table}

Now, we want to put our analyses together to determine the allowed intersections of a type $I_1$ curve of self-intersection $-1$ or $-2$.  We begin with the former case.  The residual vanishings are $(\tilde{a}, \tilde{b}, \tilde{d}) = (4,6,13)$.  We thus have the constraints
\begin{align}
\sum_P \tilde{a}_P &\leq  4 \nonumber \\
\sum_P \tilde{b}_P &\leq 6 \\
\sum_P\tilde{d}_P &\leq  13. \nonumber
\end{align}
Here, the sum runs over all points of intersection of the type $I_1$ curve $\{ z=0 \}$ with the other singular curves.  In addition, the specific form (\ref{i1form}) introduces the constraint,
\begin{equation}
\sum_P \text{ord}_P \Delta_R \leq 7.
\end{equation}
Imposing these conditions yields the maximal allowed symmetries shown in Table \ref{typei1tableb}.
\begin{table}
\begin{center}
\begin{tabular}{c|c} 
Kodaira Types & Symmetry Algebras \\\hline
$IV^{*,ns} \oplus I_3$&$\mf{f}_4 \oplus \mf{su}_3$ \\
 $I_4^{*} $&$\mf{so}(16) $\\
 $I_0^{*,ns} \oplus I_1^{*,ns}  $& $\mf{so}(7) \oplus  \mf{so}(9) $ \\
$I_7$ & $\mf{su}(7)$ \\
 $I_N^{*,ns} \oplus I_M,M +N \leq 4$ &$\mf{so}(2N+7) \oplus \mf{su}(M) $ \\
\end{tabular}
\caption{Maximal type $I_1$ intersections for $-1$ curves.}
\label{typei1tableb}
\end{center}
\end{table}

The allowed possibilities are much more constrained in the case of a type $I_1$ curve of self-intersection $-2$.  Now, the residual vanishings are $(\tilde{a}, \tilde{b}, \tilde{d}) = (0,0,2)$.  Thus, this curve can only meet a single curve of type $I_2$ or else two curves of type $I_1$.

\subsection{Type $II$}

We now turn to the case of Kodaira type $II$.  A curve $\{z=0\}$ with this fiber type has vanishing degrees $(a_{\Sigma},b_{\Sigma},d_{\Sigma})=( \geq 1,1,2)$.  Such a curve cannot collide with a curve of type $II^*$ or $III^*$ without introducing $(4, 6, 12)$ singularities.  A colliding curve with $IV^*$ fiber must be non-split since the monodromy cover for the $IV^*$ curve $\{\sigma=0\}$ of Table \ref{monodromycovertable} splits only if $z^2$ divides $\frac{g}{\sigma^4}|_{\sigma=0}$, which in turn introduces a $(4, 6, 12)$ singularity.  In the non-split case, the intersection point has $( \tilde{a}_P, \tilde{b}_P, \tilde{d}_P ) = (3,4,8)$.

No $I_{n\geq 2}^*$ curve can meet the type $II$ curve.  To see this, we write the most general forms of $f$, $g$, and $\Delta$ for such an $I_{n\geq 2}^*$ curve:
\begin{align}
f & = - \frac{1}{3} u_1^2 \sigma^2 + f_3 \sigma^3 + f_4 \sigma^4+ O(\sigma^5) \nonumber \\
g &= \frac{2}{27} u_1^3 \sigma^3 +  - \frac{1}{3} u_1 f_3 \sigma^4 + (\tilde g _5 - \frac{1}{3} u_1 f_4)\sigma^5 + O(\sigma^6) \\
\Delta &=  u_1^2 (4 u_1 \tilde{g}_5 - f_3^2) \sigma^8 + O(\sigma^8) \nonumber
\end{align}
If this curve is intersecting transversly the type $II$ curve $\{z=0\}$, we must have $z | u_1$, $z | f_3$, $z | f_4$, $z | \tilde{g}_5$.  But then, the minimal degrees of vanishing at the point of intersection are easily read off as $(f,g, \Delta)= (4,6,12)$.

An $I_1^*$ can meet the type $II$ curve, and the resulting intersection has $( \tilde{a}_P, \tilde{b}_P, \tilde{d}_P ) = (3,4,8)$.  This follows from the most general forms of $f$, $g$, and $\Delta$,
\begin{align}
f & = - \frac{1}{3} u_1^2 \sigma^2 + f_3 \sigma^3 + O(\sigma^4) \nonumber \\
g &= \frac{2}{27} u_1^3 \sigma^3 + (\tilde g _4 - \frac{1}{3} u_1 f_3)\sigma^4 + O(\sigma^5) \\
\Delta &=  4 u_1^3  \tilde{g}_4  \sigma^7 + O(\sigma^8) \nonumber
\end{align}
We see that the $\sigma^2$ term in $f$ and the $\sigma^3$ term in $g$ must vanish to order $z^2$, while the $\sigma^7$ term in $\Delta$ must vanish to order $z^3$.  However, the next-to-leading order terms need only vanish as $z$, $z$, and $z^2$, respectively, yielding $( \tilde{a}_P, \tilde{b}_P, \tilde{d}_P ) = (3,4,8)$.  This $I_1^*$ must be non-split, as the relevant term in the monodromy cover for $I_1^*$ goes as $\frac{\Delta}{\sigma^4} (\frac{f}{g})^3 |_{\sigma=0}$, which is proportional to $z$ and hence not a perfect square.  The only way to make this a perfect square is to take $\tilde g_4 \propto z^2$, but this introduces a $(4,6,12)$ point.

An $I_0^*$ can meet our type $II$ curve, and if it is non-split, the resulting point will have $( \tilde{a}_P, \tilde{b}_P, \tilde{d}_P ) = (2,3,6)$.  To consider the semi-split $\mathfrak{so}(7)$ case, we consider the $I_0^*$ monodromy cover,
\begin{equation}
\psi^3 + \frac{f}{\sigma^2}|_{\sigma=0} \psi + \frac{g}{\sigma^3}|_{\sigma=0}
\end{equation}
This cover splits only if it factorizes as
\begin{equation}
(\psi - \lambda) (\psi^2 + \lambda \psi + \mu)
\end{equation}
Comparing these two equations, we have $-\mu \lambda =  \frac{g}{\sigma^3}|_{\sigma=0}$, $\mu - \lambda^2 = \frac{f}{\sigma^2}|_{\sigma=0}$.  Now, since $z | f$, we must have $z | \mu, z| \lambda$.  This implies $z^2 | \frac{g}{\sigma^3}|_{\sigma=0}$, which means $( \tilde{a}_P, \tilde{b}_P, \tilde{d}_P ) = (2,4,6)$ for the semi-split case.

Next, we consider the fully split $\mathfrak{so}(8)$ case.  Here, the monodromy cover splits only if it factorizes as
\begin{equation}
(\psi - \alpha)(\psi-\beta)(\psi-\gamma)
\end{equation}
Since $z$ divides $f$, we must have that z divides $\alpha \beta + \alpha \gamma + \beta \gamma$.  Thus, $z$ divides at least two of $\alpha, \beta$, and $\gamma$.  If it divides all three, we get a $(4,6,12)$ singularity at the intersection point.  Matching the $\psi^2$ terms imposes $\alpha + \beta + \gamma =0$, so if we suppose $z | \alpha$, $z|\beta$, $z  \nmid \gamma$, we see that $\gamma$ must vanish.  But this implies $z | \gamma$, a contradiction.  Therefore, a fully split $I_0^*$ cannot touch a type $II$ curve.

We now consider the case of type $I_n, n\geq 1$ curves meeting a type $II$ curve.  We illustrate the case of $I_1$, which generalizes in a straightforward manner to higher $I_n$ non-split.  We then consider the split case.

The most general form of $I_1$ is as follows:
\begin{align}
f & = - \frac{1}{48} \phi^2 + f_1 \sigma + O(\sigma^2) \nonumber \\
g &= \frac{1}{864} \phi^3 + g_1 \sigma + O(\sigma^2) \\
\Delta &=  \frac{1}{192} \phi^3 (12 g_1 + \phi f_1) \sigma  + O(\sigma^3) \nonumber
\end{align}
We must have $z | \phi, z| f_1, z|g_1$.  This means that $\frac{f}{z}|_{z=0} \propto \sigma$, $\frac{g}{z}|_{z=0} \propto \sigma$, $\frac{\Delta}{z^2}|_{z=0} \propto \sigma^2$.  So, $( \tilde{a}_P, \tilde{b}_P, \tilde{d}_P ) = (1,1,2)$.

This can be straightforwardly generalized to non-split $I_n$ fibers for $n \leq 6$ and $n \geq 10$.  However, for $n=7,8,9$, the most general Tate form is not known.

For $I_n, n \geq 3$ split, we must have $z^2$ divides $\frac{f}{g}|_{\sigma=0}$.  For $n=3$, the most general form is:
\begin{align}
f & = - \frac{1}{48} \mu^2 \phi_0^2 + \frac{1}{2} \mu \phi_0 \psi_1 \sigma + f_2 \sigma^2 + f_3 \sigma^3 + O(\sigma^4) \nonumber \\
g &= \frac{1}{864} \mu^3 \phi_0^6 - \frac{1}{24} \mu^2 \phi_0^3 \psi_1 \sigma+\frac{1}{4} (\mu \psi_1^2-\frac{1}{3} \phi_0^2 f_2 ) \sigma^2 + (\tilde{g}_3 - \frac{1}{12} \mu \phi_0^2 f_3) \sigma^3+ O(\sigma^4) \\
\Delta &=  \frac{1}{16} \mu^3 \phi_0^3 (\phi_0^3 \tilde{g}_3 - \psi_1^3 - \phi_0^2 \psi_1 f_2 ) \sigma^3  + O(\sigma^4) \nonumber
\end{align}
To get a split $I_3$ fiber, then, we must have $z^2 | \mu$ or else $z | \phi_0, z| \psi_1$.  In the most optimistic case, we get an intersection point with degrees of vanishing $(f, g, \Delta) = (3,4,8)$ and $(\tilde{a}_P, \tilde{b}_P, \tilde{d}_P) = ( 2,3,6 )$.

For split $I_4$, on the other hand, the minimal form is:
\begin{align}
f & = - \frac{1}{48} \mu^2 \phi_0^4 + \frac{1}{6} \mu \phi_0^2 \phi_1 \sigma + (\hat{f_2} - \frac{1}{3} \phi_1^2) \sigma^2 + f_3 \sigma^3 + f_4 \sigma^4 + O(\sigma^5) \nonumber \\
g &= \frac{1}{864} \mu^3 \phi_0^6 - \frac{1}{72} \mu^2 \phi_0^4 \phi_1 \sigma + \frac{1}{6} (\frac{1}{3} \mu \phi_0^2 \phi_1^2-\frac{1}{2} \mu \phi_0^2 \hat{f_2} ) \sigma^2 + (-\frac{1}{3} \phi_1 \hat{f_2} + \frac{2}{27} \phi_1^3 - \frac{1}{12} \mu \phi_0^2 f_3) \sigma^3  \nonumber \\
&+ (\hat{g_4} - \frac{1}{3} \phi_1 f_3 - \frac{1}{12} \mu \phi_0^2 f_4) \sigma^4 + O(\sigma^5) \\
\Delta &=  \frac{1}{16} \mu^2 \phi_0^4 (- \hat{f_2}^2 + \mu \phi_0^2 \hat{g_4} ) \sigma^4  + O(\sigma^4) \nonumber
\end{align}
For must $I_4$, we must have $z^2 | \mu$ or $z | \phi_0$.  The na\"ive minimal degrees of vanishing of $f$, $g$, and $\Delta$ are $3$, $5$, and $10$, respectively.  This is not a Kodaira type, however, and so we conclude that $f$ must actually vanish to order 4 at the intersection point.  However, this implies $z^2 | \hat{f_2}$, which in turn raises the degrees of vanishing to $4$, $6$, and $12$.  We conclude that a split fiber of type $I_n, n\geq 4$ cannot meet a type $II$ fiber.

The only remaining fibers to consider are those of type $II$, $III$, and $IV$.  These may all intersect a type $II$ curve as expected, though the degrees of vanishing $(\tilde{a}_P,  \tilde{b}_P, \tilde{d}_P)$ for a split $IV$ fiber are increased to $(2,3,6)$ compared with the $(2,2,4)$ non-split case.  We thus have the allowed intersections with type $II$ curves shown in Table \ref{typeIItable}.
\begin{table}
\begin{center}
\begin{tabular}{c|c|c|c} 
Fiber Type & $\tilde{a}_P$ & $\tilde{b}_P$ & $\tilde{d}_P$  \\\hline
$IV^{*,ns}$ & 3&4 &8 \\
$I_1^{*,ns}$ & 3&4 &8 \\
$I_0^{*, ss}$ & 2&4 &6 \\
$I_0^{*, ns}$ & 2&3 &6 \\
$I_n^{ns}, n \geq 10$&$\lfloor \frac{n}{2} \rfloor$ & $2 \lfloor \frac{n}{2} \rfloor$ & $4 \lfloor \frac{n}{2} \rfloor$  \\
$I_9^{ns}$& $\leq 4$ & $\leq 8$ &$\leq 16$  \\
$I_8^{ns}$& $\leq 4$ &$\leq 8$ &$\leq 16$  \\
$I_7^{ns}$& $\leq 3$ &$\leq 6$ &$\leq 12$  \\
$I_6^{ns}$& 2& 4& 8\\
$I_5^{ns}$& 2& 4& 8\\
$I_4^{ns}$& 2& 4& 8\\
$I_3^{s}$ & 2 &3 & 6\\
$I_3^{ns}$ & 1 &2 &4 \\
$I_2$ & 1& 2&4 \\
$I_1$ & 1& 1& 2\\
$I_0$ & 0& 0& 0\\
$IV^{s}$ &2 &3 &6 \\ 
$IV^{ns}$ &2 &2 &4 \\ 
$III$ &1 &2 &4 \\ 
$II$ &1 &1 &2 \\ 
\end{tabular}
\caption{Orders of vanishing at type $II$ intersections.}
\label{typeIItable}
\end{center}
\end{table}

With this analysis, we may compute the collections of curves that are allowed to simultaneously meet a type $II$ curve of self-intersection $-1$ or $-2$.  In the first case, $\tilde{a} = 5$, $\tilde{b} = 7$, $\tilde{d} = 14$.  The maximal allowed symmetry algebras are then shown in Table \ref{typeIItableb}.
\begin{table}
\begin{center}
\begin{tabular}{cc} 
Kodaira Types & Symmetry Algebras \\\hline
$IV^{*,ns} \oplus I_0^{*,ns}$&$\mf{f}_4 \oplus \mf{g}_2$ \\
$I_1^{*,ns} \oplus I_0^{*,ns}$&$\mf{so}(9) \oplus \mf{g}_2$ \\
$I_0^{*,ns} \oplus I_6^{ns}$& $\mf{g}_2 \oplus \mf{sp}(3)$ \\
$I_6^{ns} \oplus I_3^{s}$& $\mf{sp}(3) \oplus \mf{su}(3)$ \\
$\leq  I_9^{ns} \oplus I_0^{*,ns}$ & $\leq  \mf{sp}(4) \oplus \mf{g}_2 $
\end{tabular}
\caption{Maximal type II intersections for $-1$ curves.}
\label{typeIItableb}
\end{center}
\end{table}
From here, we see that, indeed, the symmetry algebra is always a subalgebra of $\mathfrak{e}_8$.  One might worry that the unknown $I_8$ and $I_9$ fibers could present a problem.  However, these would at most introduce a $\mf{sp}(4) \oplus \mf{g}_2$ algebra, which is still a subalgebra of $\mf{e}_8$.

For a type $II$ curve of self-intersection $-2$, the possibilities are much simpler.  Such a curve can intersect a type $IV^{ns}$ curve, a type $III$ curve, a type $I_2$ curve, or two curves of types $I_1$ or type $II$.  The maximal gaugable symmetry allowed in F-theory is thus $\mf{su}(2)$.

\subsection{Tangencies}

Curves in the F-theory base of 6D SCFTs always intersect transversely.  In LSTs, on the other hand, curves may intersect tangentially.  In particular, we expect tangential intersections of the form:
\begin{equation}
1 \,\, || \,\, 4 \,,~~~~2 \,\, || \,\, 2
\end{equation}
Here, the parallel lines $II$ indicate a tangential intersection.

The residual order of vanishings of $f$, $g$, and $\Delta$ at tangential intersections of $I_1$ and $II$ curves can easily be determined from Tables \ref{typeI1table} and \ref{typeIItable}: they is simply double whatever appears there.  To see this, we note that a tangency between a curve $\Sigma=\{z =0\}$ and another curve $\Sigma'$ can be expressed locally in Weierstrass form by setting $\Sigma' = \{z u + \sigma^2 =0\}$.  The tangential intersection then occurs at the point $(z, \sigma ) =0$.

Consider the case of an $I_1$ fiber meeting an $I_0^*$ fiber.  When the curves $\{ z=0\}$, $\{\sigma=0\}$ intersected transversely, we had
\begin{align}
f &\sim \sigma^2 + O(z) \nonumber \\
g &\sim \sigma^3 + O(z) \\
\Delta &\sim z \sigma^6 + O(z) \\
\end{align}
From this, we read off $a_P=2$, $b_P=3$, $d_P=6$.  Now, in the tangential case, we are simply replacing $\sigma$ with $z u + \sigma^2$.  Expanding around $z=0$ to compute the residual degree of vanishing at the intersection, we therefore have
\begin{align}
f &\sim \sigma^4 + O(z) \nonumber \\
g &\sim \sigma^6 + O(z) \\
\Delta &\sim z \sigma^{12} + O(z) \\
\end{align}
As $\sigma \rightarrow 0$, we now read off the residual orders of vanishing $\tilde{a}_P=2$, $\tilde{b}_P=3$, $\tilde{d}_P=6$.  Clearly, this generalizes to arbitrary fiber type intersecting a type $I_1$ or type $II$ curve.  We immediately see that the $1 \,\, || \,\, 4$ intersection cannot occur when the fiber type of the $-1$ curve is $I_1$ or $II$, since the fiber type on any $-4$ curve is minimally $I_0^{*,s}$.  Similarly, if we set the fiber type of the $-1$ curve to be $I_0$, we see that the residual order of vanishing $\tilde{d}_P$ at the point of intersection is greater than or equal to $14$ whenever the fiber type on the $-4$ curve is $I_n^*, n \geq 1$, $IV^*$, $III^*$, or $II^*$.  This means that the only the minimal fiber type $I_0^{*,s}$ is allowed to be on a $-4$ curve tangent to a $-1$ curve, which has $(\tilde{a}_P,\tilde{b}_P,\tilde{d}_P)=(4,6,12)$.  Furthermore, this analysis also rules out the possibility of a curve of self-intersection $-5$ or below tangentially intersecting a curve of self-intersection $-1$: the former will necessarily have fiber type $IV^*$, $III^*$, or $II^*$.

With $-2$ curves, the analysis is very similar.  Once again, the residual orders of vanishing at a tangency will be twice what they were for a transverse intersection.  As a result, the $2 \,\, || \,\, 2$ intersection is only allowed if the fiber types of the two $-2$ curves are identical: either they are both type $I_0$, they are both type $I_1$, or they are both type $II$.

A configuration of two curves meeting tangentially can be deformed into a configuration of two curves meeting transversely at two distinct points.  This explains the reason for the doubling of the degrees of vanishing $\tilde{a}_P=2, \tilde{d}_P$, and $\tilde{d}_P$ at a tangency relative to a single transverse intersection.  It further allows us to relate the allowed configurations in which an unpaired $-1$ curve or $-2$ curve intersects another curve at two distinct points to the allowed configurations involving a tangency.  Namely, we see that an unpaired $-1$ curve is allowed to meet a $-4$ curve at two distinct points only if the $-1$ curve has fiber type $I_0$ and the $-4$ curve has fiber type $I_0^*$.  Similarly, an unpaired $-2$ curve can meet another $-2$ curve only if that other $-2$ curve has the same fiber type: either $I_0$, $I_1$, or $II$.

This is all compatible with the $\mathfrak{e}_8$ gauging condition for $-1$ tensors and the $\mathfrak{su}_2$ gauging condition for $-2$ tensors discussed previously, provided we correctly interpret the rules of the gauging condition for tangencies and double intersections.  If a $-4$ curve of gauge algebra $\mathfrak{so}(8)$ meets an unpaired $-1$ curve either through a tangency or a double intersection in the F-theory base, the intersection matrix is
\begin{equation}
A_{IJ} = \left( \begin{array}{cc} -4 & 2\\ 2 & -1  \end{array} \right).
\end{equation} 
Evidently, the $\mf{e}_8$ gauging condition should be modified in this situation to account for the multiplicity of the intersection between the two tensors given by the value of $A_{12}$.  To be precise, we must have
\begin{equation}
\bigoplus_{i=1}^{A_{12}} \mf{g} \subset \mf{e}_8.
\end{equation}
Here, we have $A_{12}=2$, and indeed $\mf{so}(8) \oplus \mf{so}(8) \subset \mathfrak{e}_8$, so this configuration is allowed.  However, any enhancement of the gauge algebra on the $-4$ tensor would not be allowed, as $\mf{g} \oplus \mf{g} \not\subset \mathfrak{e}_8$ for any such $\mathfrak{g}$.  By the same reasoning, an unpaired $-2$ tensor meeting a tensor carrying any nontrivial gauge algebra more than once would violate the $\mf{su}_2$ gauging condition for $-2$ tensors, since $\mf{g} \oplus \mf{g} \not\subset \mf{su}_2$ for any $\mf{g}$.  The modified gauging condition thus reproduces the F-theory result that if an unpaired $-2$ tensor meets another $-2$ tensor with multiplicity two i.e. if we have the intersection matrix
\begin{equation}
A_{IJ} = \left( \begin{array}{cc} -2 & 2\\ 2 & -2  \end{array} \right),
\end{equation} 
then both of the $-2$ tensors must be unpaired.

\section{Unusual Configurations with Unpaired Tensors}\label{sec:UNUSUAL}

Thus far, we have established that an unpaired $-1$ curve has an $\mathfrak{e}_8$ global symmetry and an unpaired $-2$ curve has an $\mathfrak{su}_2$ global symmetry visible in F-theory,\footnote{We use the phrase ``visible in F-theory" because global symmetries of 6D theories visible in F-theory do not always match field-theoretic expectations \cite{global-symmetries}. Indeed, it was argued in \cite{Heckman:2016ssk} that some 6D SCFTs feature emergent global symmetries in the IR limit, which are not visible from F-theory.  Since we are concerned with the part of the global symmetry that can be gauged by neighboring tensors in F-theory, only the global symmetry visible in F-theory is relevant to us here.} which may be gauged by adjacent tensors carrying gauge algebras.  These rules are sufficient to classify the vast majority of 6D SCFTs and LSTs involving unpaired tensors.  However, there are a handful of would-be theories which are not in violation of these rules, yet cannot be produced in F-theory \cite{Heckman:2015bfa}.  All of these involve unpaired $-2$ tensors.  Constructing these theories in another manner, or else finding a field-theoretic justification for their non-existence, is one of the most pressing issues hindering a completely field-theoretic classification of 6D SCFTs and LSTs.  In this section, we establish some preliminary results in this direction.

The first class of unusual configurations involves a single empty $-2$ curve touching a $-2$ curve with gauge algebra $\mathfrak{su}_2$.  One might expect that configurations of the form,
\begin{equation}
2 \,\, \overset{\mathfrak{su}_2}2 \,\, \overset{\mathfrak{so}_7}2
\end{equation} 
would be allowed.  However, as discussed in \cite{Ohmori:2015pia}, the global symmetry seen by the middle $-2$ tensor is $\mathfrak{g}_2$ rather than $\mathfrak{so}_7$, and a single half-hypermultiplet of $\mathfrak{su}_2$ lives at the intersection between the unpaired tensor and the middle tensor.

The second class of unusual configurations involves two or more adjacent empty $-2$ tensors.  In such cases, none of these unpaired tensors can meet a $-2$ curve carrying a gauge algebra.  For instance,
\begin{equation}
2 \,\, 2 \,\, \overset{\mathfrak{su}_2}2
\end{equation}
and 
\begin{equation}
2 \,\, 2 \,\, \overset{\mathfrak{su}_2}1.
\end{equation}
are not allowed in F-theory.  If we try to put $I_0$ on the leftmost $-2$ 
curve, the order of residual vanishing of the discriminant will be too large on this curve.  If we try to put $I_1$ or $II$ on the leftmost curve, the order of residual vanishing of the discriminant will be too large on the middle $-2$ curve.  
Evidently, the gauging condition for a $-2$ curve is more subtle than for 
a $-1$ curve: an unpaired $-2$ curve somehow gobbles up some of the 
$\mf{su}_2$ symmetry of an adjacent unpaired $-2$ 
curve. A partial explanation for this phenomenon arises 
from associating hypermultiplets to unpaired
$-2$ curves according to their intersection with the residual
discriminant, even though there is no gauge charge which would single out
those hypermultiplets in field theory.  Thus, an intersection
of two unpaired $-2$ curves would have a hypermultiplet associated
to two different tensor fields, analogous to being charged under two
distinct summands of the gauge algebra.  Indeed, an analysis of the anomaly polynomials of SCFTs with consecutive unpaired $-2$ tensors such as E-string theories \cite{Ohmori:2014pca,Ohmori:2014kda} and theories parametrized by nilpotent orbits of flavor symmetries \cite{Heckman:2016ssk} reveals the existence of a hypermultiplet at each point of intersection between unpaired $-2$ curves \cite{NRT}.

The final class of unusual configurations involve $D$-type configurations of $-2$ curves with unpaired tensors:
\begin{equation}
2 \,\, {\overset{2}{\overset{\mathfrak{su}(N_1)}{2}}} \,\, \overset{\mathfrak{su}(N_2)}{2} \,\, ...\,\, \overset{\mathfrak{su}(N_{k-1})}{2} \,\, \overset{\mathfrak{su}(N_k}{2}
\end{equation}
One might expect that we could build up a ladder of $\mathfrak{su}(N_i)$ gauge groups to arbitrarily large $N_k$, but in fact the only allowed configurations are
$$
2 \,\, \overset{2}{2} \,\, {2} \,\, ... \,\, {2}
$$
$$
2 \,\, {\overset{2}{\overset{\mathfrak{su}_2}{2}}} \,\, \overset{\mathfrak{su}_2}{2} \,\, ... \,\, \overset{\mathfrak{su}_2}{2}
$$
$$
2 \,\, {\overset{2}{\overset{\mathfrak{su}_2}{2}}} \,\, \overset{\mathfrak{su}_2}{2} \,\, ... \,\, \overset{\mathfrak{su}_2}{2} \,\, 2
$$
\begin{equation}
2 \,\, {\overset{2}{\overset{\mathfrak{su}_2}{2}}} \,\, \overset{\mathfrak{su}_3}{2} \,\, ... \,\, \overset{\mathfrak{su}_3}{2}
\label{eq:Dtype}
\end{equation}
$$
2 \,\, {\overset{2}{\overset{\mathfrak{su}_2}{2}}} \,\, \overset{\mathfrak{su}_3}{2} \,\, ...\,\, \overset{\mathfrak{su}_3}{2} \,\, \overset{\mathfrak{su}_2}{2}
$$
$$
2 \,\, {\overset{2}{\overset{\mathfrak{su}_2}{2}}} \,\, \overset{\mathfrak{su}_3}{2} \,\, ...\,\, \overset{\mathfrak{su}_3}{2} \,\, \overset{\mathfrak{su}_2}{2} \,\, 2
$$
In field theory, the plateau at $\mathfrak{su}_3$ in the last three theories can be explained by looking at the $-2$ curve with three trivalent neighbors.  The intersection point of this curve with the leftmost curve supporting a $\mathfrak{su}_3$ algebra holds a $(\bf{2+1,3})$ rather than a $(\bf{2,3})$.  Since a $-2$ curve with gauge algebra $\mathfrak{su}_3$ must hold six fundamental hypermultiplets to satisfy gauge anomaly cancellation, there are therefore only three fundamentals left that can pair up with the gauge algebra to the right.  As a result, a bifundamental $(\bf{3,4})$ of $\mathfrak{su}_3 \oplus \mathfrak{su}_4$ is not allowed, and the maximal gauge algebra that can arise is $\mathfrak{su}_3$.

Note, however, that the following theory is constructible in F-theory:
\begin{equation}
2 \,\, {\overset{\mathfrak{su}_2}{2}} \,\, \overset{\mathfrak{su}_3}{2} \,\, \overset{\mathfrak{su}_4}{2} \,\,...
\end{equation}
Here, there is only a $(\bf{2,3})$ living at the intersection of the $\mathfrak{su}_2$ curve and the  $\mathfrak{su}_3$ curve.  Evidently, the addition of the second unpaired tensor in (\ref{eq:Dtype}) changes the $(\bf{2,3})$ into a $(\bf{2+1,3})$.  We have no field-theoretic explanation for why this should be the case.

One should also wonder if theories with identical gauge algebras but distinct fiber types for unpaired $-2$ curves in F-theory flow to the same theories in the IR.  For the very simplest example of this, consider theories of a single $-2$ curve and fiber types $I_0$, $I_1$, and $II$: do these distinct F-theory models flow to the same superconformal fixed point?  Several arguments can be given in the affirmative.  The $I_0$ theory is dual to the worldvolume theory of two M5-branes and flows to the the 6d (2,0) theory of type $A_1$.  Upon compactification to 5d, this gives maximally supersymmetric Yang-Mills theory with gauge group $U(2)$.  The $I_1$ theory, on the other hand, is dual to the worldvolume theory of two M5-branes at the origin of a transverse Taub-NUT space.  When compactified on a circle, this gives rise to $\mathcal{N}=2^*$ theory in 5d, with the mass parameter of the adjoint hypermultiplet specified by the Wilson line of the Kaluza-Klein $U(1)$ around the Taub-NUT circle \cite{Haghighat:2013tka, Haghighat:2014vxa}.  If, however, we turn off this Wilson line so that the adjoint hypermultiplet becomes massless, we get maximally supersymmetric Yang-Mills theory with gauge group $U(2)$.  This implies that the 6d SCFT coming from the $I_1$ model must also be maximally supersymmetric.  Given the ADE classification of (2,0) SCFTs \cite{Cordova:2015vwa}, we conclude that this theory must be the same 6d (2,0) theory of type $A_1$ that we had for the $I_0$ model (up to free (2,0) hypermultiplets).

Additionally, the recent work \cite{Heckman:2016ssk} found that no RG flows parametrized by nilpotent elements of flavor symmetries will generate a flow from a theory with one fiber type to a theory with another: the distinction between these fiber types is invisible to these RG flows.  This gives us reason to believe that these distinct fiber types differ only by their associated numbers of uncharged hypermultiplets.  RG flows between 6D SCFTs parametrized by nilpotent elements are all examples of Higgs branch flows, which break R-symmetry but preserve Poincar\'e invariance \cite{Intriligator:2014eaa, Cordova:2015fha}.  This in turn uniquely fixes the number of free hypermultiplets appearing in the IR \cite{Heckman:2015axa,Heckman:2015ola}, which means RG flows cannot be used to distinguish between F-theory models differing only by uncharged hypermultiplets.  Aside from these free hypermultiplets, there is no apparent field-theoretic distinction between these various superconformal fixed points.

\section{Conclusions \label{sec:CONC}}

We have studied unpaired tensors from the perspective of F-theory with the aim of moving towards a completely field-theoretic classification of 6D SCFTs and LSTs.  We have verified the $\mathfrak{e}_8$ gauging condition for $-1$ curves for Kodaira fiber types $I_0$, $I_1$, and $II$, thereby finding no evidence that these distinct fiber types should give rise to distinct theories in the IR.  We have performed a similar analysis for $-2$ curves, finding a necessary but insufficient $\mathfrak{su}_2$ gauging condition.  We have subsequently listed the classes of configurations that obey these gauging conditions yet are impossible to realize in F-theory.  To claim a completely field-theoretic classification of 6D SCFTs, one must either construct these theories in field theory or else explain the reason for their nonexistence.

\section*{Acknowledgements}

We wish to thank Marco Bertolini, Jonathan Heckman, Guglielmo Lockhart, Noppadol Mekareeya, Peter Merkx, Washington Taylor, Alessandro Tomasiello, and Cumrun Vafa for useful discussions.  DRM also thanks the Institut Henri Poincar\'e for hospitality
during the final stages of this project.  The work of DRM is supported in part by National Science Foundation grant PHY-1307513 (USA) and by the Centre National de la Recherche
Scientifique (France).  The work of TR is supported by NSF grant PHY-1067976 and by the NSF GRF under DGE-1144152.


\bibliographystyle{utphys}
\bibliography{global}

\end{document}